\documentclass[ amsmath,amssymb, aps,prd,superscriptaddress,tightline,nofootinbib,numerical,floatfix,apsmath,twocolumn]{revtex4-2}
\usepackage[latin9]{inputenc}
\usepackage{amsmath}
\usepackage{graphicx}
\usepackage{appendix}
 
 \usepackage[unicode=true,
 bookmarks=false,
 breaklinks=false,pdfborder={0 0 1},backref=false,colorlinks=true]
 {hyperref}
\hypersetup{
 urlcolor=green,linkcolor=blue,citecolor=red}

\makeatletter
\usepackage{dcolumn}
\usepackage{bm}
\usepackage{braket}
\usepackage{footmisc}

\newcommand{\curly}[1]{\mathcal{#1}}

\newcommand{\eq}[1]{
    \begin{align}
        #1
    \end{align}
}
\newcommand{\del}{\partial}
\newcommand{\eps}{\varepsilon}

\makeatother

\begin{document}
\title{Gravitational Optomechanics: Photon-Matter Entanglement via Graviton Exchange }
\author{Dripto Biswas}
\affiliation{Universit\`{a} degli Studi di Torino Via Pietro Giuria, 1, I-10125 Torino,
Italy}
\author{Sougato Bose}
\affiliation{Department of Physics and Astronomy, University College London, Gower
Street, WC1E 6BT London, UK}
\author{Anupam Mazumdar}
\affiliation{Van Swinderen Institute, University of Groningen, 9747 AG, The Netherlands}
\author{Marko Toro\v{s}}
\affiliation{School of Physics and Astronomy, University of Glasgow, Glasgow, G12
8QQ, UK}
\begin{abstract}
The deflection of light in the gravitational field of the Sun is one
of the most fundamental consequences for general relativity as well
as one of its classical tests first performed by Eddington a century
ago. However, despite its center stage role in modern physics, no
experiment has tested it in an ostensibly quantum regime where both
matter and light exhibit non-classical features. This paper shows
that the interaction which gives rise to the light-bending also
induces photon-matter entanglement as long as gravity and matter are 
treated at par with quantum mechanics. The quantum light-bending
interaction within the framework of perturbative quantum gravity
highlights this point by showing that the entangled states can be generated already
with coherent states of light and matter exploiting the non-linear
coupling induced by graviton exchange. Furthermore, the quantum light-bending interaction is capable of discerning between the spin-2 and spin-0  gravitons {thus also providing a test for alternative theories of gravity at short distances and at the quantum level}.  We will conclude by estimating the order
of magnitude of the entanglement generated by employing the linear entropy. 
In particular, we find that a {half-}ring cavity of radius $0.25$~m placed around a $10$~kg mechanical oscillator operating at $150$ Hz, could be used to generate linear entropy of order unity using a petawatt laser source at optical wavelengths. {While the proposed scheme is beyond the current experimental realities it nonetheless initiates the discussion about testing the spin of the gravitational interaction at the quantum level.}
\end{abstract}
\maketitle

\noindent \textbf{\textit{Introduction.--}} 
Among the most pivotal moments in the history of general relativity was the observation
of light deflection during the solar eclipse in 1919~\citep{dyson1920ix}.
The predictions of Newtonian gravity
differed from the predictions of general relativity for the angle
of light deflection, thus providing a possibility for a definitive test
between the two theories.  Since
then, general relativity has passed numerous tests~\citep{will2014confrontation},
from laboratory experiments of gravitational redshift~\citep{pound1960apparent}
to the detection of gravitational waves~\citep{abbott2016observation}.

Nevertheless, a key question remains; whether gravity is classical or quantum and 
how would it couple to any quantum matter? The theory of quantum gravity~\citep{kiefer2006quantum}
is expected to reveal how to combine quantum mechanics with general relativity and various consequences 
for understanding problems ranging from black hole physics to the early Universe
~\citep{carlip2015quantum,oriti2009approaches}.
 However, devising a decisive test of quantum gravity, capable of falsifying the classical notion of 
 spacetime remains a daunting task 
and requires ingenious methods~\footnote{It is extremely challenging to extract any quantum feature of gravity from the cosmological perturbations~\citep{Martin:2017zxs}, and also it holds true for the primordial gravitational waves, if it were at all detectable in future~\citep{Ashoorioon:2012kh}.}.

Nevertheless, in 2017 it was shown that a quantum gravity effect can be tested
using a simple matter-wave interferometer exploiting quantum entanglement~\citep{bose2017spin}
(see also \citep{marletto2017gravitationally} for a related work).
The basic idea is that quantum gravity will induce entanglement of masses
(QGEM), which can be explained as follows; the two electrically neutral massive objects each 
placed in their spatial superposition via 
Stern Gerlach interferometry~\citep{bose2017spin,Margalit:2020qcy,Marshman:2021wyk,Zhou:2022frl} are located
close enough that the mutual quantum gravitational interaction can generate
entanglement (a {\it non-classical correlation}), but still far enough apart that all other interactions
(e.g., electromagnetic such as Casimir) are suppressed. The first such feasibility study was performed in \citep{bose2017spin}
and in \citep{vandeKamp:2020rqh,Toros:2020dbf,Wu2023,Schut:2021svd,Tilly:2021qef,Rijavec:2020qxd, carney2019tabletop,pedernales2022enhancing,pedernales2020motional,krisnanda2020observable,chevalier2020witnessing}. As the two masses cannot entangle through a local-operation-classical-communication
(LOCC)~\citep{bennett1996mixed}, as is the case for a classical
gravitational field, one must conclude that to generate entanglement
the gravitational field must be a bonafide quantum entity~\cite{bose2022mechanism,marshman2020locality}. The proposed
scheme can hence provide a model-independent test about the quantum
nature of spacetime, and by detecting entanglement one can rule out
classical models of the gravitational field~\cite{kafri2014classical,bose2017spin,marshman2020locality,bose2022mechanism,belenchia2018quantum,galley2022no,christodoulou2019possibility}.

One of the key {\it phenomenological} advantages is that the interpretation of a {\it graviton} as a quanta which mediates 
the gravitational force can be probed experimentally (e.g., a massive graviton will have different degrees of freedom and will modify the potential/force), including the properties of the spin-2 and spin-0 components of the graviton which are responsible for mediating the force. The underlying mechanism
has been analysed in detail within perturbative canonical quantum
gravity~\citep{marshman2020locality,bose2022mechanism}, in the framework
of the Arnowitt-Desse-Meissner (ADM) approach~\citep{danielson2022gravitationally},
as well as the path integral approach~\citep{christodoulou2022locally}. The QGEM protocol can also probe the quantum weak equivalence principle where both matter and gravity are treated at par~\citep{Bose:2022czr}, unlike any other existing experimental protocols where gravity is always treated classically. It can be used for quantum sensing~\citep{Toros:2020dbf} with foreseeable applications for probing new physics such as axions or fifth force, and other physics beyond the Standard Model~\citep{Barker:2022mdz}.

At the very core of this simple, but powerful result, is the idea
to test whether spacetime can mediate entanglement between the two quantum
systems. However, instead of considering two matter-waves one could
in principle also consider two massless particles such as photon pairs.
While conceptually simple, photon-photon scattering via the gravitational
interaction poses a formidable experimental challenge for a laboratory
experiment~\citep{barker1967gravitational,boccaletti1969photon}. Another option 
is to consider hybrid matter-photon setups, with a
photonic system gravitationally coupled to a heavy quantum system~\citep{whittle2021approaching},
generalising the situation of quantum field theory in a curved spacetime.
Indeed, single-photon sources have in recent years enabled
the experimental exploration of multi-mode interference and entanglement~\citep{fink2017experimental,restuccia2019photon,torovs2020revealing,torovs2022generation}
within the context of quantum field theory in curved spacetime~\citep{wald1994quantum,brunetti2015advances},
a regime which can be thought of as a stepping stone towards a quantum
theory of gravity. 

In this work we will investigate the quantum counter-part of the classical
light-deflection effect with a matter-wave and a photonic system, and
show that it leads to matter-photon entanglement as long as we assume
that the gravitational field is a quantum entity. 
We will find that the gravitationally induced quantum mechanical Hamiltonian
for the photon-matter system is a cubic interaction reminiscent of
the familiar coupling found in cavity optomechanics.
We will consider the initial state of a mechanical oscillator to be
its (coherent) ground state $\ket{0}$ and the initial state of the
photon to be a coherent state $\vert\alpha\rangle$, and estimate the
gravitationally induced matter-photon entanglement using linear entropy.

A fundamental importance of the proposed protocol, based on the  quantum light-bending effect, is that it can differentiate between spin-2 and spin-0 gravitons, thus providing a method to {distinguish between effective theories of gravity at short distances, where no classical or quantum test has been performed. It can test Perturbative quantum gravity~~\cite{Scadron:1979di,Bjerrum-Bohr:2014zsa,carney2022newton}, Brans-Dicke theory~\cite{weinberg1972gravitation}, modified gravity theories~\cite{moffat2009bending}, as well as a number of massive gravity models~\cite{panpanich2019particle}.} Discerning the spin of the mediator at a quantum level will be a crucial milestone. 

The aim will be to pin down the parameter space which will lead to ${\cal O}(1)$ position-momentum entanglement of this gravitational optomehanical system. {We will show that a {half-}ring cavity of radius $\sim 0.25$ m placed around a $10$kg system harmonically trapped at $150$ Hz ($0.1$ Hz)} could be used to generate linear entropy of order unity by using the intensities already available using petawatt (megawatt) laser sources.


\begin{figure}
    \centering
    \includegraphics[width=\columnwidth]{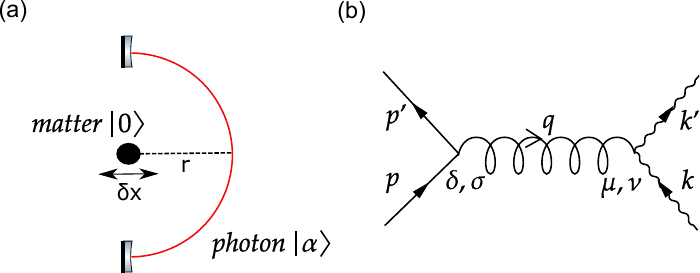}
    \caption{(a) Blueprint of the experimental scheme. The mass $m$ is harmonically trapped and prepared in the (coherent) ground state $\vert 0\rangle$. We will assume that the optical field is confined to a {half-}ring of radius $r$ (i.e., in the geometric approximation) and that the photon state is initially prepared in a coherent state $\vert \alpha \rangle$. (b) Tree-level photon-matter scattering via exchange of a graviton. Straight external lines denote the massive particle and squiggly lines denote photons.}
    \label{fig:scattdiag}
\end{figure}


\noindent \textbf{\textit{Quantum interaction.--}} 
We consider a particle of mass $m$ placed at the origin and a circular path of radius $r$ for the optical field as shown in Fig.~\ref{fig:scattdiag}(a). The gravitational interaction between the matter system and the optical field in a classical theory is 
given by~\cite{2003AmJPh..71..770B}:
\eq{
V = -\frac{2G m\omega}{r} \eps^*_{k',\nu'}\cdot \eps_{k,\nu},
\label{effpotstatres1}}
where $m$ is the mass, $\omega$ ($ \eps_{k,\nu}$) denotes the frequency (polarization vector) of the optical field, $k$ is the three-momentum, and $\nu$ denotes the polarization. In particular, the potential in Eq.~\eqref{effpotstatres1} can be computed in general relativity and gives rise to the light-bending effect ~\cite{dyson1920ix} (in the Appendix A we show how to obtain this potential in an effective field theory approach to quantum gravity). However, note that the observables are actually quantum operators~\cite{bose2022mechanism}:
\begin{align}
r & \rightarrow \hat{r}, \label{eff1}\\
\eps_{k,\nu} & \rightarrow \eps_{k,\nu} \hat{a}_{k,\nu}, \label{eff2}
\end{align}
where $\hat{r}$ is the position operator for the relative distance degree of freedom,  and $\hat{a}_{k,\nu}$ is the mode operator for the optical field. From Eqs.~\eqref{effpotstatres1}-\eqref{eff2} we then find~\cite{Scadron:1979di,Donoghue:1994dn,Bjerrum-Bohr:2014zsa}:
\eq{
\hat{V} = -\frac{2G m\omega}{\hat{r}} \eps^*_{k',\nu'}\cdot \eps_{k,\nu} \hat{a}^\dagger_{k',\nu'} \hat{a}_{k,\nu},
\label{effpotstatres2}}
which can be seen as the quantum counter-part of the light-bending interaction.

{It is important to note that Eq.~\eqref{effpotstatres2} is a non-trivial consequence of perturbative quantum gravity. The six off-shell degrees of freedom of the graviton, encoded in the spin-2 and spin-0 components~\cite{van1973ghost,marshman2020locality}, combine to give the pre-factor 2 on the right-hand side, while, for example, in an effective scalar theory of gravity (such as in N\"{o}rdstrom gravity\cite{N1,N2,N3}) the coupling would vanish altogether~\cite{zee2010quantum}. A quantitative test of the quantum light-bending interaction thus provides a conclusive test for a number of alternative theories of gravity~\cite{Scadron:1979di,Bjerrum-Bohr:2014zsa,carney2022newton,weinberg1972gravitation,
moffat2009bending,panpanich2019particle}.}

{Moreover, the quantum nature of the graviton gives rise to a quantum-interaction, with operator-valued observables, in stark contrast to the classical interaction in Eq.~\eqref{effpotstatres1}, arising from the classical theory of general relativity~\cite{Scadron:1979di}.} 
Although the steps in Eqs.~\eqref{eff1}-\eqref{eff2} seem innocuous, as it is the familiar quantization procedure commonly performed, it has non-trivial consequences for the underlying gravitational field from which the interaction arises. Indeed, the procedure of promoting the classical observables of the matter-photon system to operators, changes the nature of the gravitational interaction from classical to quantum, as we now discuss.

One can make a simple argument following~\cite{bose2022mechanism} as to why Eq.~\eqref{effpotstatres2} can no longer arise from a (real-valued) classical gravitational field.  We recall that the usual interpretation of Eq.~\eqref{effpotstatres1} is that of the energy of the gravitational field; if the right-hand side is real-valued then the energy of the gravitational field  is  real-valued and the gravitational field can have a classical description. However, as soon as we promote the classical observables to operators in Eqs.~\eqref{eff1}-\eqref{eff2} we transform the \emph{energy of the gravitational field}  to an \emph{operator valued quantity} in Eq.~\eqref{effpotstatres2}, hence requiring also an operator-valued description for the gravitational field. In particular, the right-hand side of Eq.~\eqref{effpotstatres2} contains cross-coupling terms between $\hat{r}$ and $\hat{a}^\dagger_{k',\nu'}\hat{a}_{k,\nu}$, which can generate matter-photon entanglement. Since no classical entity is capable of mediating entanglement, as formalized by the LOCC theorem, we must conclude that Eq.~\eqref{effpotstatres2} originates from bonafide quantum properties of the gravitational interaction with matter and light.



\noindent \textbf{\textit{Graviton induced optomechanical coupling.--}} 
{For concreteness} we consider the experimental configuration shown in Fig.~\ref{fig:scattdiag}(a). We assume the  photon beam at an impact parameter of $r$ with respect to the oscillator, and write
\begin{equation}
\hat{r} = {\vert (\delta \hat{x},0,0) - (r \cos\theta,r \sin\theta,0)\vert},\label{fluctuations}
\end{equation}
{where {$\delta\hat{x}$} contains the quantum fluctuations of the matter, and $\theta$ parametrizes the circular geometry of the cavity. We then expand $1/\hat{r}$ to linear order in $\delta\hat{x}$ (i.e, assuming $\delta\hat{x}\ll r$) and we integrate over the half-ring (i.e. over the angle $\theta\in[-\pi/2,\pi/2]$) to find $2\delta \hat{x}/r^2$.
From Eq.~\eqref{effpotstatres2} we thus immediately find the leading order interaction between the harmonic oscillator and the half-ring cavity}:
\eq{
\hat{V}\approx -2Gm\omega \left[\frac{1}{r} + \frac{{{2}\delta\hat{x}}}{r^2} + \curly{O}({\delta\hat{x}}^2)\right]\otimes\hat{\eps}^*(k') \cdot \hat{\eps}(k).
\label{secquant}
}
The first term in Eq.~\eqref{secquant} does not couple matter and photon degrees of freedom. Omitting also, the higher order contributions $\curly{O}({\delta\hat{x}}^2)$, we are left with the lowest order optomechanical interaction:
\eq{
\hat{V}\approx -\frac{{4}Gm\omega }{r^2} {\delta\hat{x}} \, \hat{\eps}^*(k') \cdot \hat{\eps}(k).
\label{secquant2}
}
To obtain the optomechanical coupling we now introduce the mode operators by writing:
\eq{
{\delta\hat{x}} = {\delta x_\text{zpf}}(\hat{b} + \hat{b}^\dag),\label{oscop}
}
and
\eq{
\hat{\eps}(k) = \eps_{k,\nu}\hat{a}_{k,\nu},\,\,\,
\hat{\eps}^*(k') =\eps^\dag_{k',\nu'}\hat{a}^\dag _{k',\nu'},\label{photop}
}
where {$\delta x_\text{zpf}=\sqrt{1/(2m\omega_\text{m})}$} denotes the zero-point fluctuations of the harmonic oscillator, 
$\omega_\text{m}$ is the frequency of the mechanical oscillator, and  $\hat{b}^\dag (\hat{b})$ and $\hat{a}^\dag (\hat{a})$ denote the creation (annihilation) operators of the mechanical oscillator and the photon, respectively. Here, $k,~k'$ and $\nu, \nu'$ denote the momentum and polarisation of the photon, but to simplify the analysis we will now consider the situation where $\nu\approx\nu'$ and $k\approx k'$ and will suppress the momentum-polarisation indices to ease the notation, i.e., $\hat{a}$ will denote the mode of a given polarization following the circular path shown in Fig.~\ref{fig:scattdiag} (a).

Hence by combining Eqs.~\eqref{secquant2}-\eqref{photop} we find that the interaction potential at the lowest order reduces to:
\eq{
\hat{V} =  -g_0 (\hat{b}+\hat{b}^\dag)\hat{a}^\dag \hat{a}, \label{Htotal}}
where the gravitationally-induced (single-photon) optomechanical coupling is given by
\eq{
g_0 = \frac{{4}G m \omega }{r^2 (2m\omega_\text{m})^{1/2}} \label{optomechanical}.}
We note that Eq.~\eqref{Htotal} is formally of the same form as the well-known cavity optomechanical interaction Hamiltonian~\cite{aspelmeyer2014cavity,bose1997preparation}, {and hence we can directly adapt well-established protocols to explore the coupling in Eq.~\eqref{Htotal}, i.e., gravitational optomechanics.} 

\noindent \textbf{\textit{Linear entanglement entropy.--}} 
To show the entanglement between the photon-matter subsystems,
we will calculate the linear entropy $S$ of the mechanical oscillator, given by
\begin{equation}
S=1-\text{Tr}(\rho_{\text{m}}^{2}),\label{eq:linent}
\end{equation}
where, $\rho_{\text{m}}$ is the reduced density matrix of the oscillator
subsystem, obtained by tracing out the photon degrees of freedom.

We will assume the initial state to be of the form
\eq{
\ket{\Psi(0)} = \ket{0}_{\text{m}}\otimes\ket{\alpha}_{\text{p}}
\label{instate}}
where, $\ket{0}_{\text{m}}$ and $\ket{\alpha}_{\text{p}}$ denote coherent states of the oscillator and the photon respectively. {The separable state in Eq.~\eqref{instate} can be prepared by placing the photon source sufficiently far from the mechanical oscillator, where the weak gravitationally-induced optomechanical coupling $\propto r^{-2}$ in Eq.~\eqref{optomechanical} does not have enough time to generate a measurable pre-existing entanglement. As can be estimated from Eq.~\eqref{analyticS3} below, such a condition can be readily met as the photon flight-time outside the cavity will be significantly shorter than the time spent inside the cavity.}

In the following we will use the following parameters
\begin{equation}
\mathcal{G}= \frac{g_0}{\omega_m}\,,~~t=\omega_m \tau\,, \label{kappat}
\end{equation}
where $t$ denotes the experimental time $\tau$ multiplied by $\omega_m$.
The time-evolved state using the quantum Hamiltonian in Eq.~(10) is given by~\citep{bose1997preparation},
\eq{\ket{\Psi(t)} = e^{-|\alpha|^2/2} \sum_{n=0}^\infty \frac{\alpha^n}{\sqrt{n!}} e^{i \mathcal{G}^2 n^2 (t-\sin t)}  \ket{\phi_n(t)} \otimes \ket{n}.
\label{psit}}
 Here, $\ket{n}$ is in the number basis of the photon space and \eq{\ket{\phi_n(t)} = \ket{\mathcal{G} n(1-e^{-it})}.
\label{phint}}
For computing the linear entanglement entropy of the system at a time $t$ using Eq.~(12), we first obtain the reduced density matrix $\rho_\text{m}$ of the oscillator,
\eq{
\rho&_\text{m}
= e^{-|\alpha|^2} \sum_n  \frac{|\alpha|^{2n}}{n!}\ket{\phi_n}\bra{\phi_n}, \\
\implies
\rho_\text{m}^2 &= e^{-2 |\alpha|^2} \sum_n \sum_m \frac{|\alpha|^{2(n+m)}}{n!m!}\ket{\phi_n}\braket{\phi_n|\phi_m}\bra{\phi_m}.
}
Using the number basis representation of coherent states, $\ket{\phi_m(t)} = e^{-|\phi_m(t)|^2/2} \sum_{p=0}^\infty \frac{\phi_m(t)^p}{\sqrt{p!}}\ket{p}$, we have,
\eq{
\rho_\text{m}^2 = e^{-2 |\alpha|^2} \sum_n \sum_m \frac{|\alpha|^{2(n+m)}}{n!m!}e^{-\frac{1}{2}(|\phi_m|^2 + |\phi_n|^2)} \nonumber \\
\times\left(\sum_{p=0}^\infty \frac{\phi^p_m \phi^{*p}_n}{p!}\right)\ket{\phi_n}\bra{\phi_m}.
\label{rhoO2}}
Taking the trace of Eq. (\ref{rhoO2}), we obtain,
\eq{
Tr&(\rho_\text{m}^2) = e^{-2 |\alpha|^2} \sum_n \sum_m \frac{|\alpha|^{2(n+m)}}{n!m!}e^{-(|\phi_m|^2 + |\phi_n|^2)} \nonumber \\
&\times \left(\sum_{p=0}^\infty \frac{\phi^p_m \phi^{*p}_n}{p!}\right) \left(\sum_{a=0}^\infty \frac{\phi^a_n \phi^{*a}_m}{a!}\right).
\label{finalrho2}}
Finally, using Eq. (\ref{phint}) in Eq. (\ref{finalrho2}), we write down the linear entropy of the oscillator subsystem as,
\eq{
S &= 1 - e^{-2|\alpha|^2}\Big[\sum_{n=0}^\infty \sum_{m=0}^\infty \frac{|\alpha|^{2(n+m)}}{n!m!} \nonumber \\
&\times \textrm{exp}\left(2\mathcal{G}^2 (m-n)^2 (\cos t - 1)\right)\Big].
\label{linentfinal}}
To find an approximation of Eq.~(\ref{linentfinal}) we first note that the summand is significant only for $m \sim n \sim \alpha^2$, for arbitrary $t,\mathcal{G} > 0$. Therefore, if we extend the summation domain to $m,~n \in \mathbb{Z}$, the result does not change significantly. One can write Eq.~(\ref{linentfinal}) as,
\eq{
S = 1 - \sum_{n \in \mathbb{Z}}\sum_{m \in \mathbb{Z}}\left(\frac{e^{-\Lambda}\Lambda^n}{n!}\right)g(m,n)\left(\frac{e^{-\Lambda}\Lambda^m}{m!}\right),
\label{poisum}}
where, 
\eq{
\Lambda = \vert\alpha\vert^2 \label{lambda}
}
is the mean photon number, and,
\eq{
g(m,n) = \exp(2\mathcal{G}^2(m-n)^2(\cos t - 1)).
\label{gmn}}
Further, assuming that $\Lambda$ is large, we can write the Poisson distributions in Eq. (\ref{poisum}) as continuous Gaussian distributions, and convert the summation into a double integral  over continuous variables $(m,n) \rightarrow (x,y)$ (formally, we use the Euler-Maclaurin formula and discard the Bernoulli terms, since all derivatives of the summand vanish for large $x,y$). Then, we have,
\eq{
S \approx 1 - \int_{-\infty}^{\infty} \int_{-\infty}^\infty dx~dy\,p(\Lambda,x) g(x,y) p(\Lambda,y),
\label{intS}}
where, $p(\Lambda,x) = \frac{\exp[(x-\Lambda)^2/2\Lambda]}{\sqrt{2\pi\Lambda}}$ and $g(x,y)$ is an extension of Eq. (\ref{gmn}) from $\mathbb{Z}\times \mathbb{Z}$ to all of $\mathbb{R}^2$. Eq.~(\ref{intS}) can be analytically computed to be,
\eq{
S \approx 1 - \frac{1}{\sqrt{1 + 8\mathcal{G}^2\Lambda(1-\cos t)}}.
\label{analyticS}}
Furthermore, assuming the value of $\mathcal{G}^2\Lambda\rightarrow 0$ is small we obtain a simple expression for the normalized entanglement entropy:
\eq{
S = S_\text{max} (1-\cos t).
\label{analyticS2}}
Using the definitions in Eqs.~\eqref{kappat} and \eqref{lambda} we find that the normalization has a simple expression
\eq{
S_\text{max}=4\mathcal{G}^2\Lambda= \frac{4 g_0^2\vert\alpha\vert^2}{\omega_\text{m}^2}, \label{S_0}
}
where we recall that $g_0$ is the gravitationally induced single-photon optomechanical coupling, $\vert \alpha\vert^2$ is the initial mean photon number of the coherent state, and $\omega_\text{m}$ is the frequency of the mechanical oscillator.

To enhance the generated entanglement, we will consider large values $\vert\alpha\vert\gg 1$ for the coherent state of the optical field, resulting in the light-enhanced optomechanical coupling 
\eq{
g=g_0\vert\alpha\vert. \label{leg0}
}
In practical experimental situations we will however still be limited to the regime $g < \omega_\text{m}$ such that $\mathcal{G}^2\Lambda$ remains small and the above approximations remain valid. We will thus use Eq.~\eqref{analyticS2}, which reaches its maximum value $2S_\text{max}$ when $t=\omega_\text{m} \tau = \pi$. 

We note that $S_\text{max} \propto 1/\omega_\text{m}^3$ (since from Eq.~\eqref{optomechanical} we have $g_0 \propto 1/\sqrt{\omega_\text{m}}$), which suggests to use low-frequency harmonic oscillators to increase the maximum attainable entanglement entropy. In addition, we want the experimental time to remain small in order to avoid the deleterious effect of environmental noises and decoherence. We are hence led to the  short-time regime $t=\omega_m \tau \ll 1$ such that we can use the approximation $\text{cos}{(t)}\approx 1-t^2/2$. Using Eqs.~\eqref{optomechanical}, \eqref{S_0} and \eqref{leg0}  we then find that Eq.~\eqref{analyticS2} reduces to the following simple formula for the entanglement entropy:
\eq{
S = 2 g^2 \tau^2=\frac{4G^2 m \omega^2 \vert\alpha\vert^2 \tau^2}{r^4 \omega_\text{m}}.
\label{analyticS3}}

\noindent \textbf{\textit{Parameter region.--}} 
To quantify the generated entanglement we consider the currently available state-of-the-art from two hitherto disparate fields; a high intensity light source, such as the CoReLS petawatt (PW) laser~\cite{yoon2021realization}, and a heavy, low-frequency, mechanical oscillator, such as the $10$ kg LIGO mirror~\cite{whittle2021approaching}. The optomechanical coupling, between these two systems, is induced by gravity. We now quantify the resulting entanglement entropy.

We recall that the light intensity $I$ is related to the amplitude via the formula
\eq{
\vert\alpha\vert^2=\frac{2 I}{\epsilon_0 E_\text{c}^2}, \label{intensity}
}
where $E_\text{c}$ is the electric field amplitude, and $\epsilon_0$ is the vacuum permittivity. We further suppose that the optical field is confined in a {{half-}}ring cavity of radius $r$ (see Fig.~\ref{fig:scattdiag}(a)) with the electric field amplitude given by 
\eq{
E_\text{c}=\sqrt{\frac{\omega}{2\epsilon_0 V_\text{c}}}, \label{field}
}
where {$V_\text{c}=(\pi r)(\pi \tilde{w}^2)$} is the cavity volume, and $\tilde{w}$ is the cavity waist. Combining Eqs.~\eqref{intensity} and \eqref{field} we then find that the amplitude is given by:
\eq{
\vert\alpha\vert=\sqrt{\frac{4 I  V_\text{c}}{\omega}}. \label{intensity2}
}
Let us estimate the order of magnitude of the generated entanglement for an optical field of intensity {$I=10^{13} \,\text{W}\text{cm}^{-2}$} at the optical wavelength $\lambda=1\,\mu \text{m}$ ($\omega=2\pi/\lambda$), while for the {{half-}}ring cavity we set the radius to $r=25~\text{cm}$ and the waist to $\tilde{w}=6\,\text{cm}$ {such that the total power circulating in the cavity is $\sim 1~\text{PW}$}; 
using these numbers we find from Eq.~\eqref{intensity2} the value {$\vert\alpha\vert \sim 10^{13}$}.
For the mechanical oscillator we consider the mass $m=10\,\text{kg}$ and the trap frequency  $\omega_\text{m}\sim 2\pi \times 150\,\text{Hz}$ (such a system has been recently cooled by LIGO to 11 phonons~\cite{whittle2021approaching}). In our case we will assume the trapped system to be sphere of radius $R\sim 6\,\text{cm}$ corresponding to a material of density $\rho\sim 10^{4}\, \text{kg}\text{m}^{-3}$.
From Eqs.~\eqref{optomechanical} and \eqref{leg0} we then find that the single-photon and light-enhanced optomechanical couplings are given by $g_0\sim 2\pi \times 10^{-29}~\text{Hz}$ and  {$g\sim 2\pi \times 10^{-16}~\text{Hz}$}, respectively.

To further enhance the coupling wee can envisage an experimental protocol with squeezed states. Since the gravitational light-bending interaction in Eq.~\eqref{Htotal} depends linearly on the position operator $\delta\hat{x} \propto \hat{b}+\hat{b}^\dag$ we can  enhance the interaction by delocalizing the mechanical oscillator in position using a squeezing protocol~\cite{janszky1986squeezing,lo1990squeezing,rashid2016experimental}. The resulting linear entanglement entropy in Eq.~\eqref{analyticS3}, which is a second order effect in the interaction (and hence $\propto \delta\hat{x}^2$), will thus be enhanced by a factor $e^{2\xi}$, where $\xi$ is the squeezing parameter. {In particular, we consider Eq.~\eqref{analyticS3} with the left-hand side amplified by the squeezing contribution $e^{2\xi}$ with {$\xi \sim 41$}, corresponding to the center-of-mass delocalization {$\Delta x= x_\text{zpf} \,e^\xi \sim 6~\text{cm}$ matching the radius of the mass, i.e. $\Delta x/R\sim 1$. To summarize, the experimental values form a simple hierarchy of values, i.e.,  $\tilde{w}\sim\Delta x\sim R$ and  $r=4R$.}
Using these values we find that the generated entanglement grows to unity in a time $\tau \sim 1\,\text{ms}$.}

{The proposed scheme can be also suitably modified without lowering the generated entanglement. {Depending on experimental and technical considerations we can
change simultaneously two or more parameters appearing in Eq.~\eqref{analyticS3}. For example, we can lower the laser power to $\sim 1\,\text{GW}$ by lowering the frequency to $\omega_\text{m}=2\pi \times 0.1\,\text{Hz}$ as well as increasing the experimental time to $\tau=2.5\,\text{s}$, at the cost of making the experiment more prone to low-frequency noises~\cite{Toros:2020dbf,Wu2023}.
Alternatively, we could change the photon frequency to gamma-rays, i.e., $\lambda=0.1\, \text{nm}$~\cite{watts2021photon}, obtaining a comparable entanglement entropy at $\omega_\text{m}=2\pi \times 150 \text{Hz}$ using the laser power $\sim100\,\text{GW}$.
Another option would be also to consider lighter masses, e.g., $m=100$ g corresponding to the radius $R=1\,\text{cm}$, by decreasing the mechanical frequency down to $\omega_\text{m}=2\pi \times 0.1 \text{Hz}$ and increasing the experimental time to $t=2.5\,\text{s}$, whilst reducing the total power in the cavity to $\sim 10 \,\text{TW}$. In this paragraph we have used the same geometric ratios, $\Delta x/R\sim 1$, $\tilde{w}/R\sim 1$, and $r/R=4$, as used in the previous paragraph, to ease the comparison of the parameters.} For further discussions about the available parameter space see Appendix B.

\noindent \textbf{\textit{Summary.--}} 
To summarise, this is a simple illustration of how a photon will get entangled with a quantum matter solely via quantum gravitational interaction. We have quantified the entanglement in an effective field theory approach to quantum gravity, and we have found that the linear entropy is given by the simple expression, $S = 2 g^2 \tau^2$, where $g$ is the (light-enhanced) gravitationally-induced optomechanical coupling defined in Eq.~\eqref{leg0}, and $\tau$ is the experimental time.

In a typical experimental setting, with low intensity light, the linear entanglement is extremely tiny. Nonetheless, our estimate suggests that there might be a way of probing the quantum light-bending interaction by combining heavy, low-frequency mechanical oscillators~\cite{whittle2021approaching} and  intense light sources~\cite{danson2019petawatt}. 

There will be many outstanding experimental issues we will need to understand. We will need to study the sources of decoherence {(see Appendix C \cite{schlosshauer2007decoherence,romero2011quantum,parikh2021quantum,
gunnink2022gravitational})}, devise cooling and squeezing  protocols~\cite{whittle2021approaching}, find a way for suppressing phonon vibrations~\cite{rahman2017laser}, {as well as extending the duration of petawatt laser pulses, to name a few}. Last but not least, we will need to construct a {\it witness} to read out the entanglement {(see Appendix D~\cite{duan2000inseparability,chen2020entanglement})}. 

Although it will be highly challenging to achieve all of the required experimental parameters, at the least, the current result highlights the parameter space to pursue the goal of testing the true quantum nature of gravity via graviton exchange via quantum entanglement.
{The proposed scheme can discern between the spin-2 and spin-0 character of the gravitational interaction by adapting \emph{well-established} optomechanical protocols from cavity-optomechanics~\cite{aspelmeyer2014cavity,bose1997preparation}, in the regime of short distances which remains unexplored even by classical experiments.}
The gravitational optomechanics is one of the {\it critical} outstanding tests which needs to be performed to understand quantum gravity's low energy frontier fully.

\noindent \textbf{\textit{Acknowledgements.--}} 
The authors would like to thank Daniele Faccio for discussions and for suggesting references about the petawatt laser, and Vivishek Sudhir for discussions. AM would like to thank {\it Institute for Advanced Study Princeton} for their kind hospitality and for hosting the author during the completion of this paper. AM's research is funded by the Netherlands Organisation for Science and Research (NWO) grant number 680-91-119. MT would like to acknowledge funding by the Leverhulme Trust (RPG-2020-197).

\bibliographystyle{unsrt}

\appendix

\section{Quantum photon-matter interaction via graviton\label{sec:quantum_interaction} }
{In this appendix we for completeness provide the derivation of the quantum-light bending interaction following the book \cite{Scadron:1979di}. We first obtain the graviton propagator (Sec.~\ref{gravprop}) and the vertex contributions (Secs.~\ref{vertex1} and \ref{vertex2}), which we then use to  obtain the scattering amplitude and by taking the Fourier transform the quantum light-bending potential (Sec.~\ref{SAEP}).}

\subsection{Graviton propagator} \label{gravprop}
Consider the following metric perturbation given by,
\eq{
g_{\mu\nu} = \eta_{\mu\nu} + h_{\mu\nu},
}
where, $\eta_{\mu\nu}$ is the Minkowski metric of flat spacetime, $\mu, \nu =0,1,2,3$, and we take $(-,+,,+,+)$  signature.
The Lagrangian is given by,
\eq{
\sqrt{-g}\curly{L} = \sqrt{-g}\left(-\frac{2}{\kappa^2}\curly{R} + \curly{L}_m + \curly{L}_{GF}\right),
\label{lagrangian}}
where, $\curly{R}$ denotes the Ricci scalar, $\curly{L}_m$ is the matter Lagrangian and $\curly{L}_{GF}$ is the gauge-fixing term. Expanding upto the second order in $h_{\mu\nu}$, we have,
\eq{
-\sqrt{-g}\frac{2}{\kappa^2}\curly{R} &= -\frac{2}{\kappa^2}(\del_\mu \del_\nu h^{\mu\nu} - \Box h) \nonumber \\
&+ \frac{1}{2}(\del_\alpha h_{\mu\nu}\del^\alpha e^{\mu\nu} - 2 \del^\alpha e_{\mu\alpha}\del_\beta e^{\mu\beta}),
\label{riccieq}}
where, $h = Tr(h_{\mu\nu})$ and $e_{\mu\nu} = h_{\mu\nu} - \frac{1}{2}\eta_{\mu\nu}h$. We also have,
\eq{
\curly{L}_{GF} = \zeta \del_\mu e^{\mu\nu} \del^\alpha e_{\alpha\nu},
\label{GFterm}}
where, $\zeta=1$ denotes the Harmonic gauge. Using this gauge, we can simplify the Lagrangian in (\ref{lagrangian}) and perform an integration by parts to obtain,
\eq{
\curly{L} = \frac{1}{2}h_{\mu\nu}\Box (\curly{I}^{\mu\nu\alpha\beta} - \frac{1}{2}\eta^{\mu\nu}\eta^{\alpha\beta})h_{\alpha\beta} - \frac{\kappa}{2}h^{\mu\nu}T_{\mu\nu},
}
where, $\curly{I}^{\mu\nu\alpha\beta} = \frac{1}{2}(\eta^{\mu\alpha}\eta^{\nu\beta} + \eta^{\mu\beta}\eta^{\nu\alpha})$.
The equation of motion for $h_{\mu\nu}$ is then,
\eq{
\left(\curly{I}^{\mu\nu\alpha\beta} - \frac{1}{2}\eta^{\mu\nu}\eta^{\alpha\beta}\right) \Box D_{\alpha\beta\gamma\delta} = \curly{I}^{\mu\nu}_{\gamma\delta}.
\label{eqgravprop}}
We can invert and solve (\ref{eqgravprop}) for the Feynman propagator $D^{\alpha\beta\gamma\delta}$, after taking the Fourier transform as,
\eq{
iD^{\alpha\beta\gamma\delta}(x) = \int \frac{d^q}{(2\pi)^4}\frac{i}{q^2 + i\eps}\curly{P}^{\alpha\beta\gamma\delta}, \nonumber \\
\implies \curly{P}^{\alpha\beta\gamma\delta} = \frac{1}{2}\left(\eta^{\alpha\gamma}\eta^{\beta\delta} + \eta^{\alpha\delta}\eta^{\beta\gamma}  - \eta^{\alpha\beta}\eta^{\gamma\delta} \right).
\label{spin2gravprop}}
(\ref{spin2gravprop}) gives the spin-2 and spin-0 components of the graviton propagator in the momentum space, see~\cite{van1973ghost,Scadron:1979di,marshman2020locality}, which we shall use to compute the scattering amplitude.

\subsection{Spin-0 particle-graviton vertex contribution} \label{vertex1}

The classical point particle stress tensor is given by,
\eq{T_{\delta\sigma}(x) = \sum_n \frac{p^n_\delta p^n_\sigma}{E^n}\delta^3(x - x_n(t)),
\label{pptens}}
where, the index $n$ runs over all the point particle legs in the diagram ($n = 1,2$ in our case), $E^n$ and $x_n(t)$ denotes the denotes the energy and the trajectory of the $n^{th}$ particle. 
Then, going to the momentum basis, for a 1-particle system we have,
\eq{\bra{p'}T_{\delta\sigma}(k)\ket{p} = p'_\delta p_\sigma + p_\delta p'_\sigma - \eta_{\delta\sigma}(p'\cdot p - m^2),
}
where, $p^2 = p'^2 = m^2$ is the particle mass. We have also chosen the standard Lorentz covariant normalisation of the states $\ket{p},\ket{p'}$, which absorbs a factor of $(2E)^{-1/2}$ from Eq.~\eqref{pptens}, each. The Feynman Rule for this type of vertex is therefore,
\eq{
f(p'_\delta p_\sigma + p_\sigma p'_\delta - \frac{1}{2}q^2 \eta_{\delta\sigma}),
\label{feynparticle}}
where, $\frac{1}{2}q^2 = (m^2 - p' \cdot p)$ and $f^2 =8\pi G$ is a coupling constant.


\subsection{Photon-graviton vertex contribution} \label{vertex2}
We start with the classical EM stress tensor as,
\eq{
T^{em}_{\mu\nu}(q) = \eps^{*\beta}( k') T_{\mu\nu;\beta\alpha}(q,k',k)\eps^\alpha(k), 
\label{emtens}}
where, $q = k-k'$ (see Fig. 1(b)) and $\eps^*, \eps$ denote the polarisation tensors. Then, the contribution of the photon-graviton vertex in the momentum basis is given by,
\eq{
\bra{k',\beta}&T^{em}_{\mu\nu}(q)\ket{k,\alpha} = \frac{1}{2} \Big[k'_\alpha(k_\mu \eta_{\beta \nu} + k_\nu \eta_{\beta\mu}) \nonumber\\
+ k_\beta&(k'_\mu \eta_{\alpha\nu} + k'_\nu \eta_{\alpha\mu}) - \eta_{\alpha\beta}(k'_\mu k_\nu + k_\mu k'_\nu) \nonumber \\ 
+~ \eta_{\mu\nu}(k' \cdot &k \eta_{\alpha\beta} - k_\beta k'_\alpha) - k'\cdot k (\eta_{\mu\alpha} \eta_{\nu\beta} + \eta_{\mu\beta} \eta_{\nu\alpha})\Big].
\label{emmom}}
The Feynman Rule corresponding to this interaction is just,
\eq{
f T^{em}_{\mu\nu;\beta\alpha}(q,k',k)
\label{feynem}}

Henceforth, we shall also make the small grazing angle approximation, wherein $q^2 = (k-k')^2 \approx 0 \implies k\cdot k' \approx 0$. Therefore, the vertex contributions in (\ref{feynparticle}) and (\ref{feynem}) simplify to,
\eq{
f(p'_\delta p_\sigma + p_\sigma p'_\delta),
\label{feynparticleapprox}}
and,
\eq{
\frac{f}{2} \Big[k'_\alpha(k_\mu \eta_{\beta \nu} + k_\nu \eta_{\beta\mu}) + k_\beta(k'_\mu \eta_{\alpha\nu} + k'_\nu \eta_{\alpha\mu}) \nonumber \\
- \eta_{\alpha\beta}(k'_\mu k_\nu + k_\mu k'_\nu) - \eta_{\mu\nu} k_\beta k'_\alpha)\Big],
\label{feynphotonapprox}}
respectively.

\subsection{Scattering Amplitude and Effective Potential}\label{SAEP}
Using the results of (\ref{spin2gravprop}), (\ref{feynparticleapprox}) and (\ref{feynphotonapprox}), the amplitude for the scattering process in Fig. 1(b) is,
\eq{
S^{cov}_{fi}& = (-i)^2 4 f^2 p'_\delta p_\sigma \frac{i \curly{P}^{\delta\sigma\mu\nu}}{q^2 + i \eps} \Big[k'_\alpha k_\mu \eta_{\beta\nu} + k_\beta k'_\mu \eta_{\alpha\nu} \nonumber \\
&- \eta_{\alpha\beta}k'_\mu k_\nu - \frac{1}{2}\eta_{\mu\nu}k_\beta k'_\alpha\Big]\eps^{\beta*}(k')\eps^\alpha(k) \delta^4(P_i),
\label{ampl}
}
The particle four-momenta, before and after scattering are denoted by $p$ and $p'$ respectively, while the photon momenta are labeled by $k, k'$. Note, that we have made the $q^2 \approx 0$ approximation only in the numerator, while the denominator is left untouched. The factor of $4$ in (\ref{ampl}) arises from the two possible momentum configurations of the photon and the point particle being contracted by the symmetric $\curly{P}^{\delta\sigma\mu\nu}$ propagator. 

Eq.~(\ref{ampl}) can be further simplified by inserting the expression for $\curly{P}^{\delta\sigma\mu\nu}$ from (\ref{spin2gravprop}) as,
\eq{
S^{cov}_{fi}& = (-i)^2 i f^2 p'_\delta p_\sigma \frac{1}{q^2 + i \eps}\Bigg[k'_\alpha k^\delta \delta^\sigma_\beta + k_\beta k'^\delta \delta^\sigma_\alpha \nonumber \\
&- \eta_{\alpha\beta}k'^\delta k_\sigma - \frac{1}{2}k_\beta k'_\alpha \eta^{\sigma\delta} + k'_\alpha k^\sigma \delta^\delta_\beta + k_\beta k'^\sigma \delta^\delta_\alpha \nonumber \\ 
- \eta_{\alpha\beta}k'^\sigma k^\delta &- \frac{1}{2}\eta^{\delta\sigma}k_\beta k'_\alpha - \eta^{\delta\sigma}\Big(k'_\alpha k_\beta + k_\beta k'_\alpha - \eta_{\alpha\beta}(k'\cdot k) \nonumber \\
- &2\eta^{\delta\sigma}k_\beta k'_\alpha\Big)\Bigg]\eps^{\beta*}(k')\eps^\alpha(k) \\
= -& 2i f^2 p'_\delta p_\sigma \left(\frac{-2 k'^\sigma k^\delta}{q^2 + i\eps}\right)\eps^*(k')\cdot\eps(k),
\label{ampleval}
}
where, in writing the last line, we have used $k^\mu \eps_\mu = 0$. 

From Eq.~\eqref{ampleval} we calculate the effective potential for a massive particle of mass $m$ and a photon of frequency $\omega$ using the  small momentum transfer limit $t \rightarrow -\vec q^2$ as,
\eq{
V(\vec r) &= \frac{1}{4m\omega} \int S^{cov}_{fi}(\vec q) e^{i\vec q \cdot \vec r} \frac{d^3 q}{(2\pi)^3} \nonumber \\
\simeq \frac{1}{4m\omega}\int&\frac{d^3 q}{(2\pi)^3} 4f^2 \frac{(k \cdot p)(k' \cdot p')}{- q^2} \eps^*(k')\cdot \eps(k) e^{i\vec q \cdot  \vec r} \nonumber \\
=~ &\frac{f^2(k\cdot p)(k' \cdot p')}{m\omega r}\eps^*(k')\cdot \eps(k),
\label{effpot}}
where $r=\vert \vec r\vert$. Finally, on taking the static limit wherein, $p_\mu = p'_\mu = M g_{\mu 0}, k_\mu \equiv (\omega, k), k'_{\mu} \equiv (\omega',  k')$ (with $\omega \rightarrow \omega'$), we get from Eq.~\eqref{effpot},
\eq{
V(r) =& -\frac{1}{4m\omega}\frac{8 \pi G (m \omega)^2}{\pi r}\eps^*(k')\cdot \eps(k) \nonumber \\
&= -\frac{2G m\omega}{r} \eps^*(k')\cdot \eps(k).
\label{effpotstatres}}
Eq.~(\ref{effpotstatres}) has been derived in previous literature (e.g., see Refs.~\cite{Scadron:1979di,Donoghue:1994dn,Bjerrum-Bohr:2014zsa}).

As discussed in the main text, it is important to note that in the context of quantum field theory, $r$ is an operator valued quantity along with the photon polarization vector $\eps(k)$, a non-trivial result emerging from perturbative quantum gravity. To highlight this crucial point we rewrite Eq.~\eqref{effpotstatres} as:
\eq{
\hat{V} = -\frac{2G m\omega}{\hat{r}} \hat{\eps}^*(k')\cdot \hat{\eps}(k).
\label{effpotstatres-1}}
Starting from this operator-valued potential, which is induced by the quantized gravitational field, we will now show that it leads to a nonlinear optomechanical interaction and matter-photon entanglement.

\section{Compact setup parameters} \label{optimal}
In this section we discuss how to optimize the parameters of the experimental scheme without reducing the generated entanglement entropy. We will first identify the length-scales of the problem and then discuss the available range of masses and mechanical frequencies (Sec.~\ref{mass1}), as well as the possible values for the laser power and the laser frequency (Sec.~\ref{laser1}).

We first combine Eqs.~(17) and (20) from the main text and find that the (gravitationally-induced) entanglement is given by:
\begin{equation}
S =  \frac{16 G^2 m  I  V_\text{c} \tau^2 e^{2\xi}}{r^4  }\frac{\omega}{\omega_\text{m}}.\label{Sapp}
\end{equation}
We recall that $V_\text{c}=(2\pi r)(\pi \tilde{w}^2)$ is the cavity volume ($\tilde{w}$ is the cavity waist),  $r$ is the distance between the cavity and the mechanical oscillator, and $m$ is the mass of the oscillator. Assuming a spherical particle, we can define the particle radius $R=(3m/(4\pi\rho))^{1/3}$, where $\rho$ denotes the density of the material. We thus see that the the generated entanglement entropy $S$ depends on three length scales: $r$, $\tilde{w}$, and $R$. 

Let us find an optimized geometric configuration by fixing the ratios $R/r$ and $\tilde{w}/r$ (and rewrite the expression for $S$ as a function of the length-scale $R$). We note that the generated entanglement in Eq.~\eqref{Sapp} scales as $r^{-4}$ and hence we want to the optical cavity to be close to the mechanical oscillator. We set the first ratio for concreteness to 
$\mathfrak{q}\equiv R/r =1/4$ 
to allow the mechanical oscillator to freely move (the minimal geometric limit would be $R/r\sim1$). Furthermore, we recall that we require $\tilde{w} \ll r$ to work within the geometric approximation; to avoid introducing a new parameter we thus set the second ratio for simplicity to $\tilde{w}/r = \mathfrak{q}=1/4$. We then find that Eq.~\eqref{Sapp} simplifies to:
\begin{equation}
S =  \frac{128\pi^3 \mathfrak{q}^3}{ 3 }\frac{G^2 I\omega \rho \tau^2  R^2 e^{2\xi}}{ \omega_\text{m}}\propto  \Delta x^2 m^{5/3},\label{Sapp2}
\end{equation}
where we have introduced the delocalization parameter $\Delta x=(2m\omega_\text{m})^{-1/2}\, e^\xi$.

\subsection{Mass and mechanical frequency}\label{mass1}
Let us assume for concreteness that the bulk of the mechanical oscillator is composed of Bismuth and estimate the generated entanglement using Eq.~\eqref{Sapp2}. Here we will consider an optical field of intensity $I=10^{13} \,\text{W}\text{cm}^{-2}$ at the optical wavelength $\lambda=1\,\mu m$ ($\omega=2\pi/\lambda$) and optimize the mass and mechanical frequency (see Sec.~\ref{laser1} below to see how to relax the requirements of the optical field).  We find that for $m=10\text{kg}$ (radius $R=6\,\text{cm}$ and density $\rho=9.747\text{g cm}^{-3}$) we can generate unit entanglement already with a mechanical frequency of $\omega_\text{m}\sim2\pi \times 150 \, \text{Hz}$ and squeezing parameter $\xi=41$, which corresponds to a spread of the center of mass wavefunction of $\Delta x= x_\text{zpf} \,e^\xi \sim 6~\text{cm}$. One could be tempted also to consider a mechanical oscillator with a smaller mass $m$. If we set $m=100\text{g}$ (radius $R=1~\text{cm}$ and density $\rho=9.747\text{g cm}^{-3}$) we can generate unit entanglement using the mechanical frequency $\omega_\text{m}=2\pi \times 1 \, \text{Hz}$ and setting the 
delocalization to $\Delta x= x_\text{zpf} \,e^\xi \sim  R \sim 1~\text{cm}$ (corresponding to a squeezing parameter $\xi=35$). 

\subsection{Laser power and photon frequency}\label{laser1}
Instead of a petawatt laser source it is desirable to use lower laser powers. A possible approach is to lower the mechanical frequency and to increase the experimental time, whilst keeping the ratio $\omega_\text{m} \tau$ of the same order of magnitude (in the previous section $\omega_\text{m}=2\pi \times 150 \, \text{Hz}$ and $\tau=1\,\text{ms}$). In particular, From Eq.~(17) in the main text, and using for simplicity $\tau =\pi/(2 \omega_\text{m})$, we readily find 
\begin{equation}
S = \frac{\pi^3 G^2 m \omega^2 \vert\alpha\vert^2 e^{2\xi}}{r^4 \omega_\text{m}^3}, \label{myE}
\end{equation}
where we note the favourable scaling $\omega_\text{m}^{-3}$. By lowering the frequency to $\omega_\text{m}=0.1\,\text{Hz}$ as well as increasing the experimental time to $\tau=2.5\,\text{s}$ we can lower the laser power to the gigawatt range. 

An alternative possibility might be to change the photon frequency $\omega$ from optical frequencies to gamma-rays. From Eq.~\eqref{myE} we note that the generated entanglement entropy scales as $\omega$ (note that $\vert\alpha\vert^2\propto 1/\omega$) and hence by changing the wavelength from $\lambda=1\,\mu\text{m}$ (optical) down to $\lambda=0.1\,n\text{m}$ (gamma-rays) we can reduce the laser power by 4 orders of magnitude, resulting in 100 megawatts of laser power. The technical details for manipulating X-rays or gamma-rays goes beyond this work, but we note that recently entanglement has been witnessed with gamma-rays~\cite{watts2021photon}, suggesting further exploration of this option.

\section{Robustness analysis}\label{RA}

In this section we discuss the deleterious effects arising from the interaction with the environment. For the experimental regime suggested in the main text we quantify the strength of the competing effects and discuss how to mitigate environmental decoherence.

The entanglement between the system and the experimental equipment (and hence the resulting environmental decoherence) can be suppressed by a suitable hierarchy of masses, distances, and frequencies, with the stringiest requirement coming from (electromagnetic) optomechanical entanglement (Sec.~\ref{oMent}), while gravitationally-induced entanglement with the experimental equipment will remain negligible (Sec.~\ref{gie}). We also find that stochastic noise from gravitons is negligible for the considered experimental parameters (Sec.~\ref{stochastic}). Finally, we quantify the experimental requirements to mitigate decoherence induced by the residual gas molecules and black-body radiation (Sec.~\ref{emdec}).

\subsection{Electromagnetically-induced entanglement with experimental equipment}\label{oMent}
The photons interact electromagnetically with a number of optical elements (such as the cavity mirrors), which could result in unwanted electromagnetically-induced entanglement, precluding the observation of the entanglement from the light-bending interaction. It is thus important to estimate the (electromagnetic) optomechanical couplings between the  optical elements and the photons, and find the parameter-regime where their effects can be mitigated. 

We can model the experimental apparatus (such as an optical element) as another harmonic oscillator of mass $M$, and frequency $\omega_M$. The standard cavity-optomechanical single-photon coupling is given by~\cite{aspelmeyer2014cavity}
\begin{equation}
g_0^{(\text{eM})}= \frac{\omega_\text{cav}}{L}x^{(M)}_\text{zpf}, \label{goM}
\end{equation}
where $\omega_\text{cav}$ is the cavity resonance-frequency, $L$ is the cavity length, and $x^{(M)}_\text{zpf}=(2M\omega_M)^{-1/2}$ is the zero-point motion associated to the optical element. 

To suppress the effect of the (electromagnetic) optomechanical coupling we require that the corresponding maximal entanglement entropy, which we will label by $S_\text{max}^{(eM)}$, should be negligible. The maximal entanglement entropy derived in Eq.~\eqref{S_0} applies also for the electromagnetic case by making the formal replacements:
 $g \rightarrow g_0^{(\text{eM})}$ and $\omega_\text{m} \rightarrow \omega_M$. In particular, we require that $S_\text{max}^{(eM)}$  should be much smaller than the gravitationally-induced entanglement  (assumed to be of order unity), i.e.
\begin{equation}
S_\text{max}^{(\text{eM})}=\left(\frac{g^{(\text{eM})}}{\omega_M} \right)^2 \ll 1, \label{Scondition}
\end{equation}
where we have introduced the light-enhanced coupling $g^{(\text{eM})}\equiv g_0^{(\text{eM})}\vert \alpha \vert$ (we recall that $\vert\alpha\rangle$ is the coherent state of the optical field). 

From  Eq.~\eqref{goM} and \eqref{Scondition} we find that the (electromagnetic) optomechanically-induced entanglement scales as $S_\text{max}^{(\text{eM})} \propto M^{-1} \omega_M^{-3}$. In other words, its magnitude can be suppressed by considering heavy optical elements (large $M$) which are strongly confined (stiff mechanical frequency $\omega_M$). Using Eq.~\eqref{goM}, and setting $\omega_\text{cav}\sim \omega$, $L=2\pi r$, we find that the condition in Eq.~\eqref{Scondition} is satisfied  for $M=1\,\text{kg}$ and $\omega_M=2\pi \times 100\, \text{GHz}$ (with the other values set to the values from the main text; $r=0.25\,\text{m}$, $\lambda=1\,\mu m$ ($\omega=2\pi / \lambda$),  $m=10\,\text{kg}$, and $\omega_\text{m}=2\pi \times 150\, \text{Hz}$).

\subsection{Gravitationally-induced entanglement with experimental equipment} \label{gie}
The entanglement between the photons and the experimental apparatus could potentially arise also through the quantum light-bending interaction. We can estimate the maximum generated entanglement using a similar analysis to the one discussed in the previous Sec.~\ref{oMent}.

The (gravitational) single-photon coupling is given by:
\begin{equation}
g_0^{\text{(gM)}} 
= \frac{2G M \omega }{r_M^2}x_\text{zpf}^\text{(M)}, \label{optomechanicalE}
\end{equation}
where $M$ ($\omega_\text{M}$) is the mass (mechanical frequency) of the experimental apparatus, $x_\text{zpf}^\text{(M)}=(2M\omega_M)^{-1/2}$ is the zero-point motion, and $r_M$ is the characteristic distance from the photons in the cavity (see Eq.~(11) in the main text). We then introduce the light-enhanced coupling $g^{\text{(gM)}}=g_0^{\text{(gM)}} \vert\alpha\vert$, where $\vert \alpha \rangle$ is the coherent state of the optical field, and require that the maximum entanglement with the experimental apparatus, $S_\text{max}^{(\text{gM})}$, should be negligible. Specifically, we require that
\begin{equation}
S_\text{max}^{(\text{gM})}=\left(\frac{g^{(\text{gM})}}{\omega_M} \right)^2 \ll  1. \label{Scondition2}
\end{equation}
Since $S_\text{max}^{(\text{gM})} \propto M \omega_M^{-3}r_M^{-4}$ we can suppress the generated entanglement using light experimental equipment (small $M$), confined in a high-frequency harmonic trap (stiff $\omega_M$), that is located far from the optical field in the cavity (large $r_M$). The characteristic distance from any localized optical element located close to the {half-}ring cavity will be approximately equal to the radius of the cavity $r_M\approx r$. The mass of the equipment, $M$, can be larger than the mass of the probe system, $m$, but the former will be trapped in a much stiffer trap, i.e., $\omega_M \gg \omega_m$. We thus have that the generated entanglement is negligible, i.e., $S_\text{max}^{(\text{gM})} \ll 1$.

One could also generate gravitationally-induced entanglement between the mass, $m$, and the experimental equipment of mass, $M$. From the Newtonian potential we find that the dominant coupling for such interaction is given by~\cite{bose2022mechanism,gunnink2022gravitational}:
\begin{equation}
g_0^{\text{(gmM)}} 
= \frac{G \sqrt{m M}}{r_{mM}^3\sqrt{\omega_m \omega_M}}, \label{gmM}
\end{equation}
where $r_{mM}$ is the distance between the two masses. We can then compute the maximum entanglement entropy, $S_\text{max}^{(\text{gmM})}$, and require that it is negligble, i.e.,
\begin{equation}
S_\text{max}^{(\text{gmM})}=\frac{(g^{(\text{gmM})} e^\xi)^2}{\omega_m\omega_M} \ll 1, \label{Scondition3}
\end{equation}
where we have included also the squeezing factor $e^\xi$ arising from the mass $m$. We note that $S_\text{max}^{(\text{gmM})} \propto M \omega_M^{-3}$ and hence we ideally want light experimental equipment (small $M$) and stiff harmonic traps (large $\omega_M$). Let us suppose that the distance between the two masses is approximately $r_{mM}\approx r$. We then find that the condition in Eq.~\eqref{Scondition3} is readily satisfied using the parameters found in Sec.~\ref{oMent} ($M=1\text{kg}$, $\omega_M=2\pi \times 100\, \text{GHz}$, $r=0.25\text{m}$, $m=10\text{kg}$, and $\omega_\text{m}=2\pi \times 150\, \text{Hz}$). 

\subsection{Stochastic noise from gravitons} \label{stochastic}
For completeness we compare the strength of the light-bending interaction with the stochastic graviton noise investigated in \cite{parikh2021quantum}. The adimensional stochastic graviton noise spectrum, assuming a graviton vacuum state (see below for a discussion about squeezed graviton states), is given by $S_{NN}(\tilde{\omega})=4 G \tilde{\omega}$, where $\tilde{\omega}$ denotes the frequency (we use the  $\tilde{\omega}$  notation to avoid confusion with the photon frequency $\omega$). This latter stochastic noise induces position fluctuations of the distance, $r$, between the mass and the photons in the cavity. Specifically, the position power spectral density is given by $S_{xx}(\tilde{\omega})\equiv r^2S_{NN}(\tilde{\omega})$, and using the relation between force and acceleration, $F(\tilde{\omega})=m \tilde{\omega}^2 x(\tilde{\omega})$, we  find the force-noise spectrum:
\begin{equation}
S_{FF}(\tilde{\omega})=4 G r^2 m^2 \tilde{\omega}^5.\label{Sff}
\end{equation}
We are interested in the dephasing rate given by~\cite{Toros:2020dbf}:
\begin{equation}
\Gamma = S_{FF}(\omega_m) \Delta x^2,\label{deltax}
\end{equation}
where $S_{FF}$ is evaluated at the mechanical frequency $\omega_m$, and $\Delta x=x_\text{zpf}\,e^{\xi}$ is the delocalization of the mechanical oscillator ($x_\text{zpf}=(2m\omega_m)^{-1/2}$ is the zero-point motion, and $\xi$ is the squeezing parameter).

Using the numbers quoted in the main text ($m=10\,\text{kg}$, $r=0.25\,\text{m}$ and $\omega_m=2\pi \times 150\, \text{Hz}$), we find that the dephasing rate $\Gamma$ is smaller by about 8 orders of magnitude compared to the (gravitationally-induced) light-enhanced coupling $g_\text{eff}=g e^\xi$ (where we have included the mechanical squeezing factor $e^\xi$ as we have done in Eq.~\eqref{deltax}). However, the stochastic graviton noise can be augmented by considering the noise originating from  a different graviton state. For example, for a squeezed graviton state we find that $\Gamma$ in Eq.~\eqref{Sff} gets rescaled by $(\text{cosh}(2 \xi^\text{(sg)}))^{1/2}$, where $\xi^\text{(sg)}$ is the graviton squeezing parameter~\cite{parikh2021quantum}. We find that the augmented dephasing rate $\Gamma$ becomes comparable to the (gravitationally-induced) light-enhanced coupling $g_\text{eff}=g e^\xi$ when we set the stochastic graviton squeezing parameter to $\xi^\text{(sg)}\sim 18$.

\begin{figure}
    \centering
    \includegraphics[width=1\columnwidth]{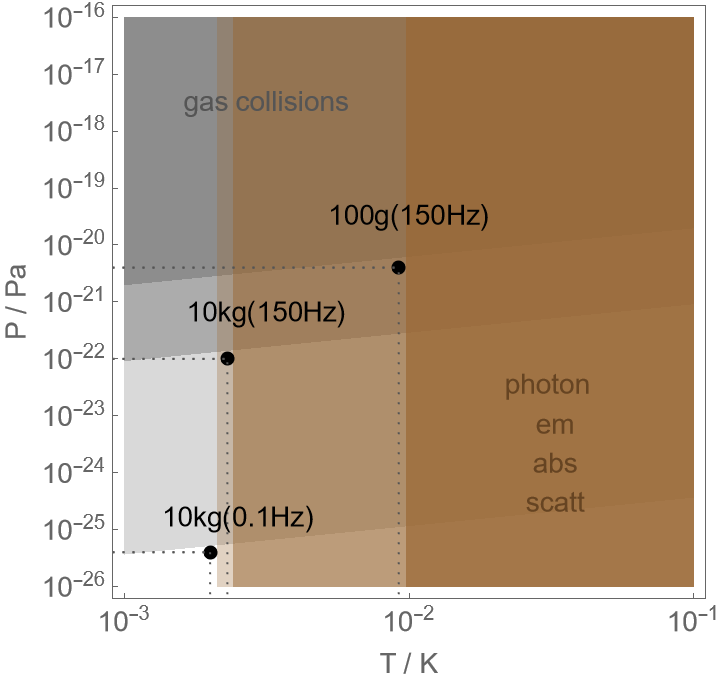}
    \caption{Plot showing the required temperature, T, and pressure, P, to achieve a coherence time of $t_\text{coh}$ as defined in Eq.~\eqref{tcoh}. $T$ denotes the temperature of the external photon bath, the temperature of the residual gas, as well as the internal temperature of the trapped particle (all three denoted by the same symbol, and assumed equal for simplicity). The brown (gray) shaded regions correspond to the parameter space excluded by considering emission, absorption, and scattering of environmental photons (collisions of residual gas particles). The shade corresponds to the mass (mechanical frequency): $m=10\,\text{kg}$ ($\omega_\text{m}=0.1\,\text{Hz}$), $m=10\,\text{kg}$ ($\omega_\text{m}=150\,\text{Hz}$), $m=100\,\text{g}$ ($\omega_\text{m}=150\,\text{Hz}$) delocalized by $\Delta x\sim 6\,\text{cm}$, $\Delta x\sim 6\,\text{cm}$, $\Delta x\sim 1\,\text{cm}$, respectively (from lighter to darker shade). The required coherence time is set to $\tau=1\,\text{ms}$ ($\tau=2.5\,\text{s}$) when the harmonic frequency is $\omega_m=2\pi \times 150\,\text{Hz}$ ($\omega_m=2\pi \times 0.1\,\text{Hz}$).  For concreteness we consider a spherical Bismuth particle with density $\rho=9.747\text{g cm}^{-3}$ (radius $R=(3m/(4\pi\rho))^{1/3}$), and dielectric constant  $\epsilon=-19.489+2.0864\text{i}$.}
    \label{fig:decplot}
\end{figure}

\subsection{Residual gas particles and environmental photons} \label{emdec} 
The mechanical oscillator of mass $m$ is subject to decoherence due to collisions with residual gas particles (`coll' label) as well as due to electromagnetic emission (`em' label), absorption (`abs' label) and scattering (`scatt' label) of environmental photons.  The decoherence rates in the long-wavelength (`lw' label)  limit are given by~\citep{schlosshauer2007decoherence,romero2011quantum,Sinha_2022}
\begin{alignat}{2}
\gamma_{\text{coll}}^\text{(lw)} & =   \frac{8 \sqrt{2 \pi } R^2\zeta (3)}{3 \zeta (3/2)}
\frac{m_g^{1/2} P}{k_B T} (k_B T)^{3/2}\Delta x^2\,,\\
\gamma_\text{em,abs,scatt}^\text{(lw)} & =\frac{8!\zeta(9)8 R^{6}}{9\pi}\left(k_{B}T\right)^{9}\text{Re}\left(\frac{\epsilon-1}{\epsilon+2}\right)^{2}\Delta x^2\,,
\end{alignat}
and the decoherence rates for the saturated short-wavelength (`sw' label) limit are given by
\begin{alignat}{2}
\gamma_\text{coll}^\text{(sw)} & =\frac{16\pi n_{V}R^{2}}{3}\sqrt{\frac{2\pi k_{B}T}{m_{g}}}\,,\label{swr1}\\
\gamma_\text{em,abs,scatt}^\text{(sw)} & =2\pi^{-1} R^2 T^3 \zeta (3) k_B^3\,,
\label{swr2}
\end{alignat}
where $R$ is the radius of the harmonically trapped system, $\zeta(\,\cdot\,)$ is the Riemann zeta function, $m_g$ is the characteristic mass of a gas particle, $P$ is the pressure of the gas, $k_B$ is the Boltzmann constant, $\epsilon$ is the dielectric constant, and $n_{V}$ is the number
density of the residual gas. $T$ denotes the temperature of the gas particles, the temperature of the external photon bath as well as the internal temperature of the trapped particle (all three denoted by the same symbol, and assumed equal for simplicity). $\Delta x$ denotes the length-scale of our system, in our case it corresponds to the zero-point motion, $r_\text{zpf}$, augmented by the squeezing factor, $e^{\xi}$,i.e., we have $\Delta x=x_\text{zpf}\, e^{\xi}$. 

We will interpolate between the short and asymptotic long wavelength regimes using the min function, i.e., $\gamma_{j} \equiv \text{min}(\gamma_{j}^\text{sw},\gamma_{j}^\text{lw})$ (for $j=\text{coll}$ and $j=\text{abs, em, scatt}$). The total decoherence rate can be then computed as 
\begin{equation}
 t_\text{coh}^{-1}\equiv \gamma_\text{coll} +\gamma_\text{abs, em, scatt}, \label{tcoh}
\end{equation}
where $ t_\text{coh}$ is the available coherence time.
In Fig.~\ref{fig:decplot} we have identified the optimal temperature and pressure to mitigate the effect of environmental decoherence of the mechanical oscillator.

\section{Adaptation of protocols from cavity-optomechanics}

The analysis in the main text showed that the quantum light-bending interaction reduces to the familiar optomechanical interaction:
\begin{equation}
\hat{V} =  g_0 (\hat{b}+\hat{b}^\dag)\hat{a}^\dag \hat{a}, \label{HtotalA}
\end{equation}
where $\hat{a}$ ($\hat{b}$) is the optical (mechanical) mode, and $g_0$ is the (gravitationally-induced) single-photon coupling (see Eqs.~(10) and (11) in the main text). Importantly, the form of Eq.~\eqref{HtotalA} matches the familiar interaction found in (electromagnetically-induced) cavity-optomechanics -- we can thus directly apply existing protocols from quantum optomechanics~\cite{aspelmeyer2014cavity}. In this section we outline a possible method to experimentally measure the entanglement arising from the gravitational interaction (from the quantum light-bending interaction discussed in the main text), by adapting the experimental protocol from~\cite{chen2020entanglement}.

\begin{figure}[t!]
    \centering
    \vspace{0.5cm}
    \includegraphics[width=0.8\columnwidth]{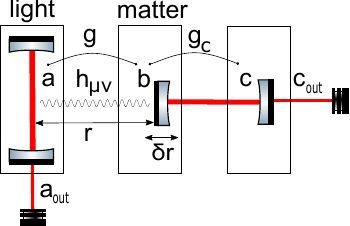}
    \caption{Measurement scheme for probing gravitationally-induced entanglement arising from the quantum light-bending interaction. The matter system is trapped harmonically with the position fluctuations denoted by $\delta \hat{x}=\delta x_{\text{zpf}}(\hat{b}+\hat{b}^{\dagger})$, where $\hat{b}$ is the mechanical mode, and $\delta x_{\text{zpf}}$ denote the zero-point fluctuations. The optical modes $\hat{a}$ and the mechanical mode $\hat{b}$ interact solely via gravity with the (gravitationally-induced) light-enhanced optomechanical coupling denoted by $g$. The mechanical mode $\hat{b}$ is also coupled to the optical mode $\hat{c}$ via the (electromagnetically-induced) light-enhanced optomechanical coupling $g_{c}$, which allows control and read-out of the mechanical motion. To ascertain the gravitationally induced entanglement between $\hat{a}$ and $\hat{b}$ we can measure EPR-type variables constructed from the output modes $a_{\text{out}}$ and $c_{\text{out}}$ and use the DGCZ criterion~\cite{duan2000inseparability, chen2020entanglement}.}
    \label{fig:witness}
\end{figure}

We first linearise the interaction by writing $\hat{a}\rightarrow \alpha+ \delta \hat{a}$, where we have introduced the mean value $\alpha=\langle \hat{a} \rangle$, and we can assume without loss of generality that $\alpha$ is real-valued. The interaction from Eq.~\eqref{HtotalA} thus simplifies to:
\begin{equation}
\hat{V}_\text{grav} =  g (\hat{b}+\hat{b}^\dag)(\delta\hat{a}^\dag +\delta \hat{a}), \label{HtotalL1}
\end{equation}
where we have introduced the (gravitationally-induced) light-enhanced optomechanical coupling $g=g_0 \vert \alpha\vert$ (see Eq.~(14) in the main text). In addition, we suppose that the mechanical oscillator is coupled also to another optical mode, $\hat{c}$, via an (electromagnetically-induced) light-enhanced coupling $g_c$, with the interaction given by:
\begin{equation}
\hat{V}_\text{em} =  g_c (\hat{b}+\hat{b}^\dag)(\delta\hat{c}^\dag +\delta \hat{c}), \label{HtotalL2}
\end{equation}
where $\delta\hat{c}$ denote fluctuations around the mean value $\langle \hat{c} \rangle$.

The schematic depiction of the setup is given in Fig.~\ref{fig:witness}. The quantum light-bending interaction entangles the optical cavity mode, $\hat{a}$, and the mechanical oscillator mode, $\hat{b}$,  the latter measured using an auxiliary optical field, $\hat{c}$. By detecting entanglement between the out optical modes $\hat{a}_\text{out}$ and  $\hat{c}_\text{out}$ we can infer the gravitationally induced entanglement between the mechanical oscillator mode $\hat{b}$ and the optical field mode $\hat{a}$.

We suppose that both cavities are driven, and that the output fields are subject to continuous measurement, analogous to the situation discussed in ~\cite{chen2020entanglement}. Assuming the initial state of the system is Gaussian, and working in the linearized regime of Eqs.~\eqref{HtotalL1} and \eqref{HtotalL2}, the final state will in general be an entangled Gaussian state. From the output modes $\hat{a}_\text{out}$ and $\hat{c}_\text{out}$ one can then construct EPR-type variables and use the Duan-Giedke-Cirac-Zoller (DGCZ) criterion to ascertain entanglement~\cite{duan2000inseparability}, or perform full Gaussian homodyne tomography to reconstruct the state of the system. Any detected entanglement between the two out modes can thus be used to ascertain the quantum nature of the light-bending interaction.

\bibliography{main}
\end{document}